\documentclass[conference]{IEEEtran}
\IEEEoverridecommandlockouts
\usepackage{cite}
\usepackage{amsmath,amssymb,amsfonts}
\usepackage{algorithmic}
\usepackage{graphicx}
\usepackage{textcomp}
\usepackage{xcolor}
\usepackage{hyperref}
\usepackage{graphicx,makecell}
\def\BibTeX{{\rm B\kern-.05em{\sc i\kern-.025em b}\kern-.08em
    T\kern-.1667em\lower.7ex\hbox{E}\kern-.125emX}}
\begin{document}

\title{Leveraging Inexpensive Lab Tools \\ for Hands-on RF Systems Course
\thanks{Funded by the Department of Electrical and Computer Engineering at Carnegie Mellon University.}
}

\author{\IEEEauthorblockN{B. Joel Gonzalez, Tom J. Zajdel, L. Richard Carley}
\IEEEauthorblockA{Department of Electrical and Computer Engineering, Carnegie Mellon University, Pittsburgh, PA, USA}
Corresponding author: B. Joel Gonzalez (bgonzale@andrew.cmu.edu)
}
\maketitle

\begin{abstract}
As wireless technologies continue to develop and reshape our world, teaching students the fundamental building blocks of RF engineering is paramount. Relatively inexpensive lab equipment now makes teaching concepts such as the Internet-of-Things, beam steering, and MIMO accessible. A new course for students in Electrical and Computer Engineering at Carnegie Mellon University aims to utilize affordable hardware to teach core concepts in RF systems design. This is accomplished through six laboratory exercises, culminating in a live demonstration of RF beam steering technology for students. The course can be run for less than \$9,000 in one-time costs and is well-received by students as evidenced by survey data, with lab resources available in an open repository.
\end{abstract}


\section{Introduction}

\subsection{Motivation}

Hands-on learning is a crucial experience for students learning communications and RF engineering to students in electrical engineering. In this era of the Internet-of-Things, wireless sensor networks are growing in size, complexity, and ubiquity. Lectures remain a core part of the curriculum to share the theory and background necessary to work in these fields. However, having students work directly with real hardware presents them with an opportunity to learn skills and concepts that will greatly benefit them in future research and industry work \cite{caverly2021use} \cite{carminati2023soak}. Unfortunately, the equipment and tools required for students to learn and apply concepts in RF engineering have historically been prohibitively expensive. High-performance vector network analyzers (VNAs), software-defined radios (SDRs), and related simulation software tools can cost thousands of dollars each, which may prohibit their use in an educational setting and limit the number of students who can use these tools \cite{derickson2021nanovna} \cite{9596897}. This course aims to take advantage of recent developments in RF measurement hardware and software-defined radio that make these tools much more affordable for universities and schools to use in educational curricula.

Our course is titled ``Board-Level RF Systems for the Internet-of-Things". The context for this course is the future in which every device wants to communicate with the world (i.e., the Internet-of-Things). It aims to teach students how radios work at a systems level, and the relevant theory about RF waves to understand signal interference, multipath propagation, and other key aspects of RF communications systems. This allows students to approach next-generation technologies, namely massive MIMO (multiple input multiple output) networks and their transformative potential, with an understanding of their fundamental components \cite{larsson2017}. The following is a list of topics that students learn in the course through lectures and labs:

\begin{itemize}
    \item Basics of Electromagnetics and RF Waves
    \item RF Transmission Lines and Impedance Matching
    \item RF Analysis Tools (Vector Network Analyzers)
    \item RF Antennas (Monopoles, Dipoles, Patch, etc.)
    \item RF Transceiver Architectures
    \item Modulation Techniques (Analog, Digital)
    \item Information Theory and Digital Communications
    \item Software-Defined Radios (Point-to-Point)
    \item Multi-Input Multi-Output RF Antennas
    \item MIMO Theory and Communications
\end{itemize}
 
\subsection{Course Design and Audience}

The 12-unit course is offered to undergraduate students (specifically third and fourth years) as 18-427 and graduate students as 18-727 at Carnegie Mellon University. The course is primarily taken by students in the ECE department. The course requires prerequisite knowledge of circuits and signal processing at the undergraduate level, in addition to basic programming experience in Python or MATLAB. Beyond these requirements, the course material should be approachable to any student who seeks to expand their understanding of RF hardware at the systems-level.

This paper provides an overview of the laboratory exercises that students perform, along with a summary of the costs to operate the course for a semester. We conclude with an evaluation of the course via student survey data from the Fall 2025 semester, illustrating how students received each hands-on laboratory exercise.

\section{Laboratory Exercises}
This section will describe the laboratory exercises that students complete, which are designed to complement lectures and to provide them with direct access to RF hardware that may otherwise be difficult to obtain and work with. Each lab session is two hours long, which provides ample time for students to complete the exercises using the equipment in groups. Table I presents a summary of the lab exercises that students complete in the course; these labs are available in an online repository \cite{github2026}.

\begin{table}[h!]
    \centering
\caption{Description of course lab exercises.}
\label{tab:my_label}
    \begin{tabular}{|c|c|}\hline
         Lab Exercise& Lab Description\\\hline
         Lab 1& RF network analysis of transmission lines\\\hline
         Lab 2& Design and fabrication of patch antennas\\\hline
         Lab 3& Characterization of patch antennas\\\hline
         Lab 4& Demodulation of RF signals\\\hline
         Lab 5& Modeling and simulation of antenna arrays\\\hline
         Lab 6& Demonstration of RF beam steering\\ \hline
    \end{tabular}    
\end{table}

\subsection{Lab 1: VNAs \& Transmission Lines} This lab uses a VNA to demonstrate RF network measurement. Students are provided with a set of PCBs manufactured by the course staff: A segment of microstrip line, a patch antenna, and a T-line junction, as shown in Fig. 1. Students use the LibreVNA to characterize these PCBs. The LibreVNA is an open-source FGPA-based 2-port vector network analyzer that can characterize RF systems from 100 kHz to 6 GHz \cite{librevna2025}, shown in Fig. 2. In comparison to the inexpensive NanoVNA popular with hobbyists which is limited to 1 GHz, the LibreVNA offers superior performance with a larger operating frequency range relevant to IoT devices \cite{10058814}, which is demonstrated as students sweep over frequencies ranging from 500 MHz up to 3 GHz in the lab exercise. In lecture, students learn the fundamentals of RF wave propagation and transmission line theory, which they then visualize through this lab as they take impedance measurements of each PCB. Students are asked to observe the phenomena of a T-junction splitter and to think critically about the impedance presented by the junction. By working with these physical tools, students gain a first-hand exposure to concepts such as reflections (i.e. S-parameters) and impedance matching networks (i.e. quarter-wave transformers).

\begin{figure}[h!]
    \centering
    \includegraphics[width=0.7\linewidth]{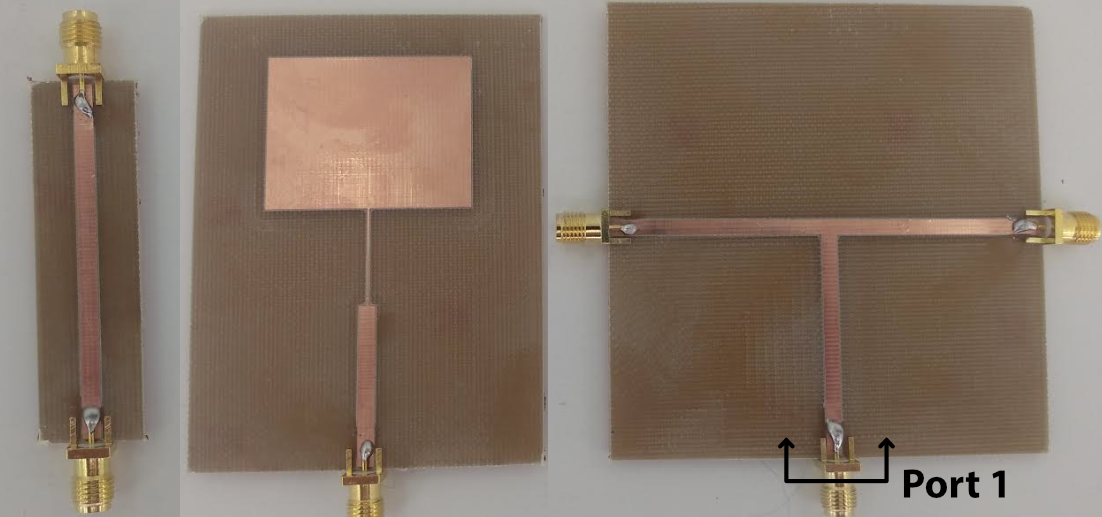}
    \caption{A set of a microstrip, patch antenna, and t-line junction PCBs.}
    \label{fig:enter-label}
\end{figure}

\begin{figure}[h!]
    \centering
    \includegraphics[width=0.4\linewidth]{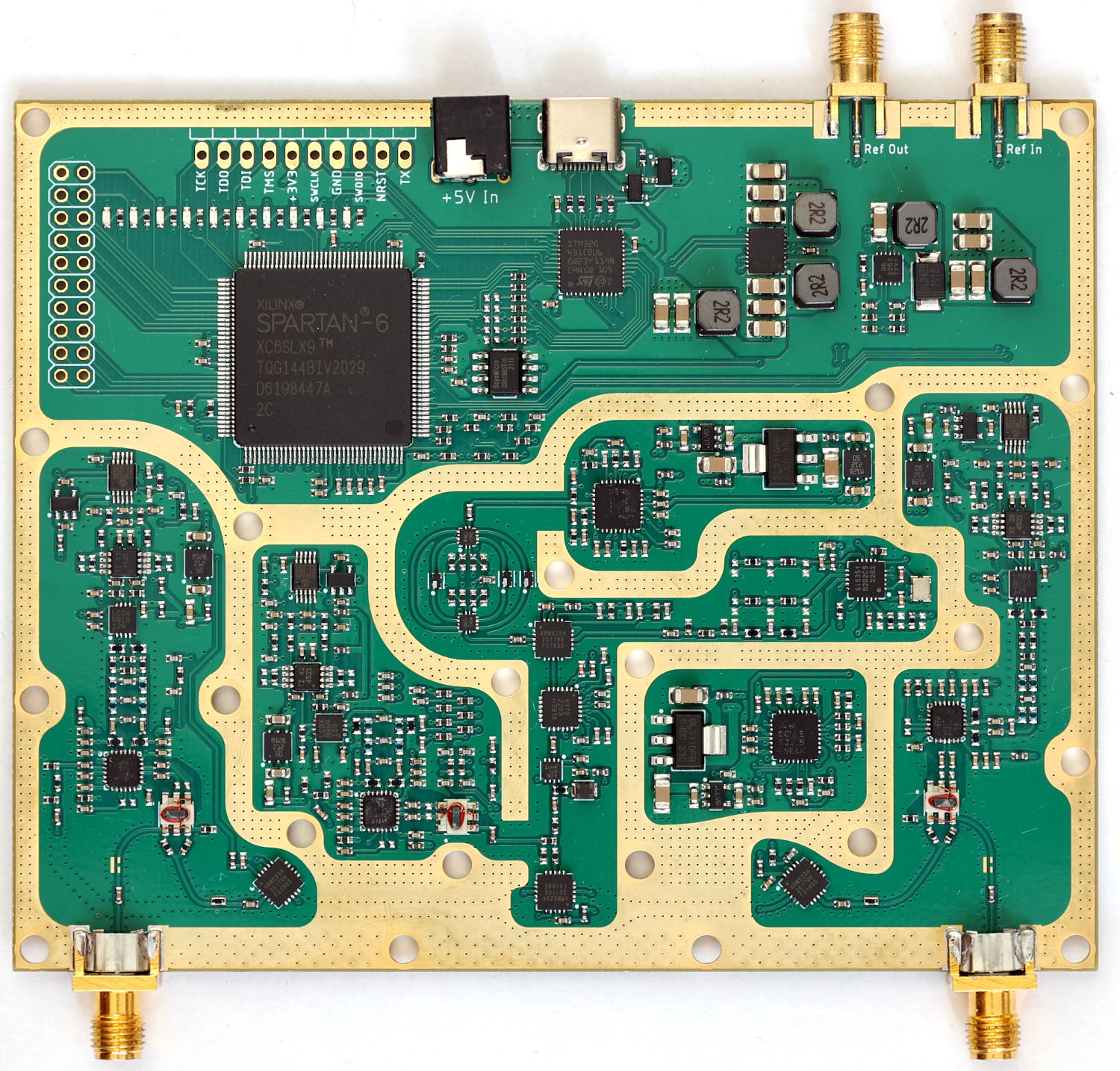}
    \caption{The LibreVNA, an open-source vector network analyzer \cite{librevna2025}.}
    \label{fig:enter-label}
\end{figure}

\subsection{Lab 2: Antenna Design \& Simulation} This lab has students investigate the design and fabrication of antennas using MATLAB's Antenna Toolbox. This package allows students to quickly and easily prototype, analyze, and visualize antenna elements without needing to learn a specific FEA design software \cite{matlab2025}. Students gain experience using the toolbox to design their own dipole and patch antennas, one of which is shown in Fig. 3, to resonate at specific frequencies. Primed by a lecture that describes the operation of patch antennas, students iterate upon their designs in simulation until they achieve an antenna that reaches the desired specification. These antennas are then fabricated by the course staff to be tested in the following lab. In past semesters, the course staff used an LPKF S63 circuit board plotter (LPKF Laser \& Electronics, Tualatin, OR) to mill out the patch antennas from a copper-clad FR4 substrate. Nowadays, instructors may submit the PCB files for inexpensive fabrication through an external vendor such as JLCPCB or PCBWay with fast turnaround time.

\begin{figure}[h!]
    \centering
    \includegraphics[width=0.7\linewidth]{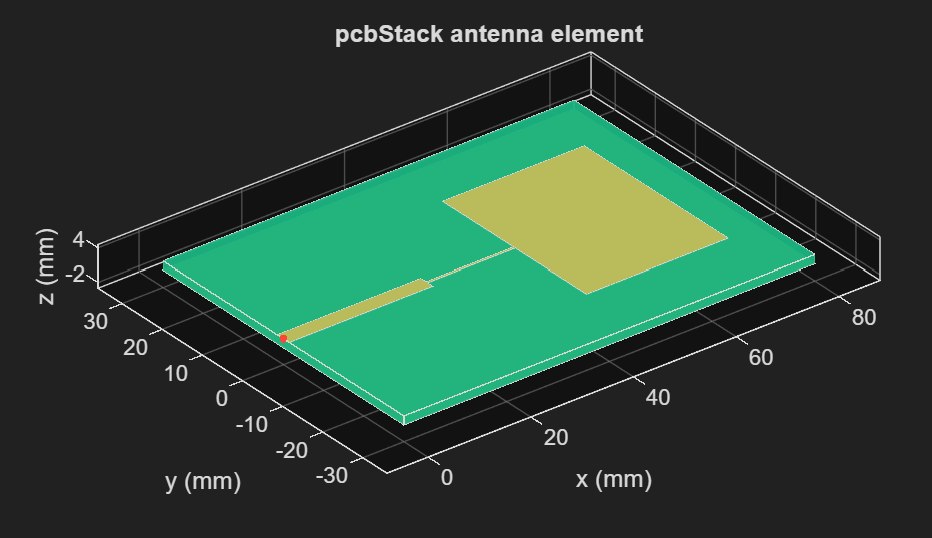}
    \caption{A patch antenna designed using MATLAB's Antenna Toolbox.}
    \label{fig:enter-label}
\end{figure}

\subsection{Lab 3: Patch Antenna Characterization} This lab allows students to see the results of their efforts from the previous lab as they test their patch antenna. Using an anechoic chamber housed at Carnegie Mellon University, students attach their patch antenna to a fixture that faces a receive antenna. As Fig. 4 illustrates, both the device-under-test (the patch antenna) and the receive antenna are connected the LibreVNA. Students map out the azimuth radiation of their antennas by providing an impulse signal to the transmit antenna and measuring the resultant forward gain ($S_{21}$). The fixture that the antenna is attached to is motorized, allowing for a Python script to automate the rotation and measurements of the antenna. An anechoic chamber is not necessary for this lab; it can be performed outdoors in a spacious environment to minimize multipath propagation from reflections \cite{osha1990}.

\begin{figure}[h!]
    \centering
    \includegraphics[width=0.7\linewidth]{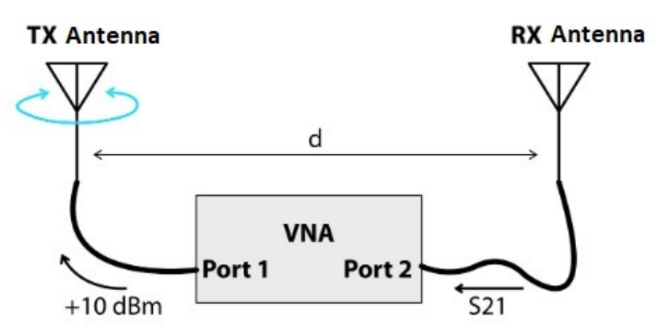}
    \caption{Antenna characterization setup diagram. The TX antenna is the device under test (DUT), while the RX antenna is a horn antenna in the anechoic chamber pointed towards the DUT. Alternatively, in an outdoors setting, the RX antenna can also be a half-wavelength dipole antenna. }
    \label{fig:enter-label}
\end{figure}

\subsection{Lab 4: Software-Defined Radios} This lab presents a shift in the course from an emphasis on RF network analysis and the physical layer toward a software-defined approach to RF systems. Students learn about the fundamental building blocks of radios through lecture, with an emphasis on the theory underpinning software-defined radios. Then, through this lab, students gain experience programming an SDR using GNU Radio, a free open-source GUI tool to interface with SDRs. Students use an RTL-SDR (RTL-SDR Blog, USA), shown in Fig. 5, which is an inexpensive receive-only SDR that can receive signals up to 1.75 GHz \cite{rtlsdr2025}. This lab exercise has students receive and demodulate live radio signals from nearby broadcast FM stations. In addition, a HackRF One (Great Scott Gadgets, Lakewood, CO), a low-cost RX/TX SDR also shown in Fig. 5, broadcasts QPSK signals in the classroom \cite{hackrf2025} so that students may demodulate them with their SDR receivers using a Costas loop \cite{4051948}. Subsequent analysis questions encourage students to think about the mathematics underlying these modulation schemes.


\begin{figure}[h!]
    \centering
    \includegraphics[width=0.4\linewidth]{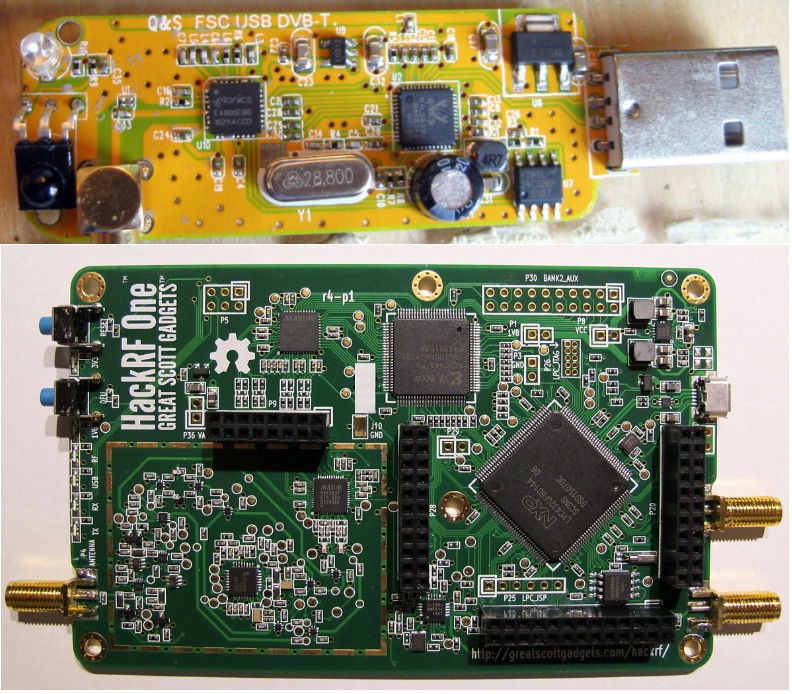}
    \caption{The RTL-SDR (top) and HackRF (bottom). Each student receives one of the RTL-SDRs, which is used to receive and demodulate RF signals. The HackRF is used to broadcast QPSK signals for students to demodulate.}
    \label{fig:enter-label}
\end{figure}

\subsection{Lab 5: Antenna Array Simulation} This lab revisits the MATLAB Antenna Toolbox to explore antenna arrays, in anticipation of the final beam-steering lab of the course. Students design and simulate dipole arrays and patch antenna arrays, as shown in Fig. 6, to better understand the directionality of these arrays. Drawing from a mathematical derivation presented in lecture, students calculate the necessary phase shifts to apply to each antenna element in order to steer the beams in a particular direction. Through this exercise, students gain an intuition for the underlying mechanisms of beam steering, as well as further developing their skills in using antenna simulation tools.

\begin{figure}[h!]
    \centering
    \includegraphics[width=0.7\linewidth]{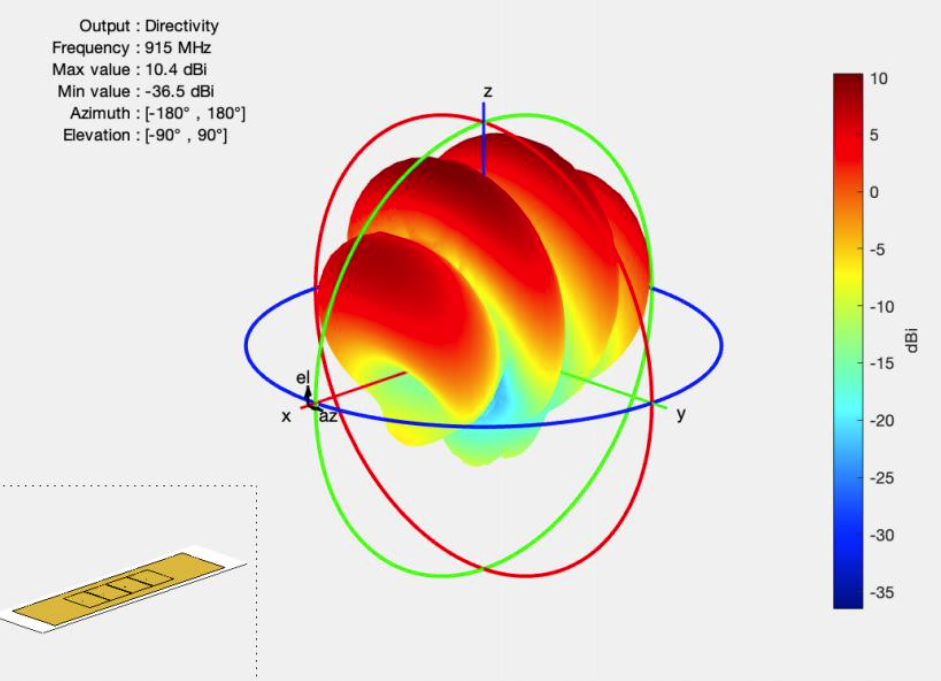}
    \caption{Antenna array simulation using MATLAB's Antenna Array.}
    \label{fig:enter-label}
\end{figure}

\subsection{Lab 6: Beam Steering \& LoRa} The final lab of the course is a live demonstration of receive beam steering using the KrakenSDR (KrakenRF Inc, Chicago, IL), shown in Fig. 7. This radio system features the same R820T2 chip as used on the RTL-SDR, but with a synchronized clock to ensure that the system is temporally coherent. Students are provided with LoRa Featherwing microcontrollers (Adafruit, Brooklyn, NY), also shown in Fig. 7, and antennas which can transmit in the 433MHz or 915MHz bands. By arranging the whip receive antennas in a linear array and having students sit in a semi-circle facing the antenna array, lab groups can take turns transmitting a LoRa signal to the receive station. As Fig. 8 illustrates, the KrakenSDR receives and processes the signal using the MUSIC algorithm on a Raspberry Pi 4 \cite{schmidt1986multiple}, which then transmits the data to a GUI that shows the direction-of-arrival of the signals \cite{kraken2025}. Students can clearly see the lobes that they are generating from their microcontroller's signal. Analysis questions then ask students to reflect upon the experimental setup, including what improvements could be made to the experiment, and how the various system components work. Through this lab, students are well-equipped to set up and perform their own experiments in wireless design and testing, pursuing their curiosity with hands-on work.

\begin{figure}[h!]
    \centering
    \includegraphics[width=0.7\linewidth]{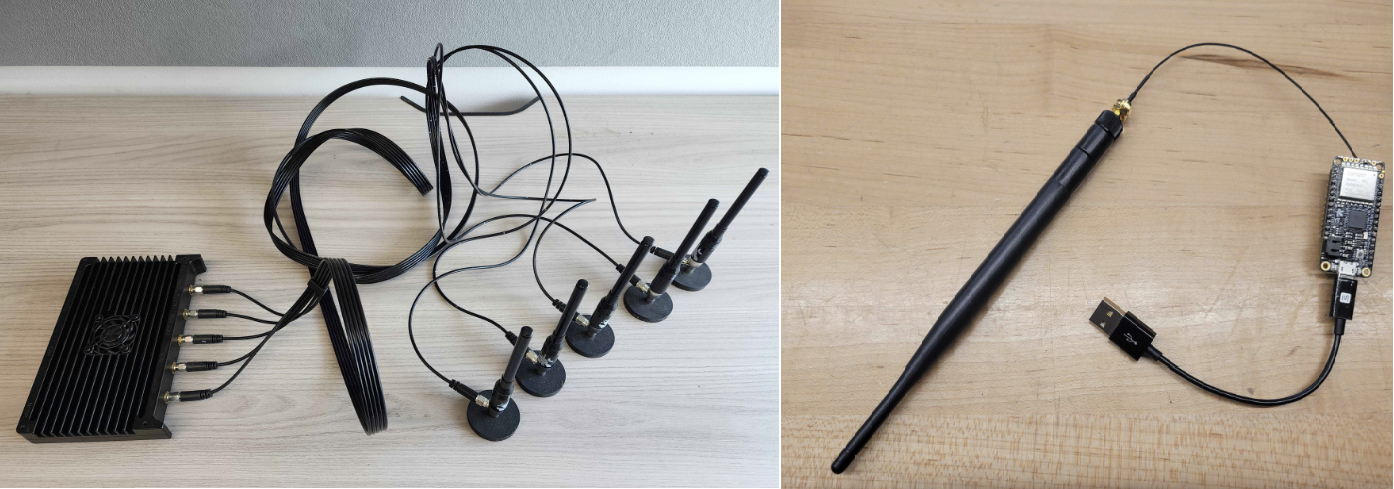}
    \caption{The KrakenSDR (left), a receive-only beam steering board, and the LoRa Featherwing microcontroller (right) that students use to transmit to the receive antenna array.}
    \label{fig:enter-label}
\end{figure}

\begin{figure}[h!]
    \centering
    \includegraphics[width=0.6\linewidth]{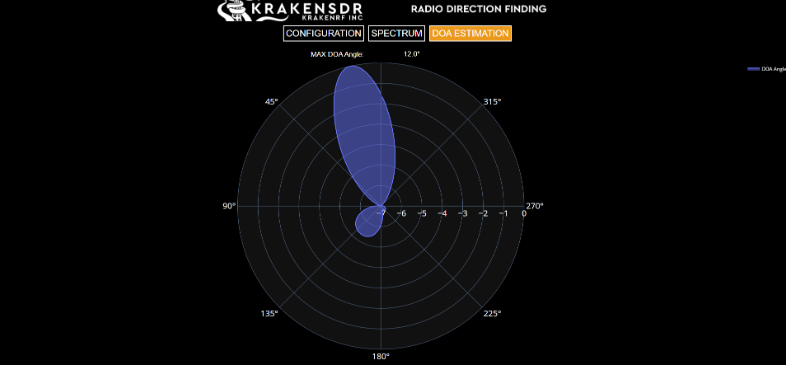}
    \caption{Diagram showing the direction-of-arrival of a LoRa signal in the classroom.}
    \label{fig:enter-label}
\end{figure}


\begin{table*}[htbp]\tiny
    \centering
\caption{Lab survey data, showing mean (scale of 1 to 7) and standard deviation in parenthesis.}
\label{tab:my_label}
    \begin{tabular}{|c|c|c|c|c|c|c|}\hline
         Question & \makecell{Lab 1 \\ ($n$=12)} & \makecell{Lab 2 \\ ($n$=14)} & \makecell{Lab 3 \\ ($n$=13)} & \makecell{Lab 4 \\ ($n$=14)} & \makecell{Lab 5 \\ ($n$=12)} & \makecell{Lab 6 \\ ($n$=12)}
\\\hline
         \makecell{This lab experience was interesting to me.} & 6.33 (0.78) & 6.42 (0.67) & 6.50 (0.67) & 6.25 (0.97) & 6.33 (0.89) & 6.25 (0.87)
\\\hline
         \makecell{This lab was easy to complete.} & 5.58 (1.08) & 5.25 (1.36) & 6.08 (1.16) & 5.83 (1.11) & 6.08 (1.00) & 6.66 (0.65)
\\\hline
        \makecell{This lab was valuable to me in \\ developing my understanding of what VNAs are used for.} & 6.08 (0.79) &  &  &  &  & 
\\\hline
        \makecell{This lab was useful in teaching me \\ the fundamental equations governing transmission line theory.} & 5.83 (1.03) &  &  &  &  &
\\\hline
        \makecell{This lab was valuable in teaching me \\ how to design a variety of antennas.} &  & 6.00 (1.04) &  &  &  &
\\\hline
        \makecell{This lab was useful in explaining \\ the process that goes into fabricating patch antennas.} &  & 5.83 (1.34) &  &  &  &
\\\hline
        \makecell{This lab helped improved my understanding \\ of how an antenna is characterized.} &  &  & 6.42 (0.79) &  &  &
\\\hline
        \makecell{This lab helped elucidate \\ the fundamental equations of antenna theory.} &  &  & 6.08 (1.38) &  &  &
\\\hline
        \makecell{This lab helped me understand \\ how to use a software-defined radio.} &  &  &  & 5.92 (1.38) &  &
\\\hline
        \makecell{This lab strengthened my understanding of \\ how demodulation works.} &  &  &  & 5.83 (1.64) &  &
\\\hline
        \makecell{This lab helped me understand \\ how to simulate beam-steering phenomena.} &  &  &  &  & 6.25 (0.86) &
\\\hline
        \makecell{This lab improved my understanding of \\ the equations that govern the physics of beam-steering.} &  &  &  &  & 6.08 (1.51) &
\\\hline
        \makecell{This lab improved my understanding of \\ how a beam-steering SDR works.} &  &  &  &  &  & 6.33 (0.78)
\\\hline
        \makecell{This lab helped me understand \\ how I could set up my own beam-steering experiments in the future.} &  &  &  &  &  & 6.08 (1.16)
\\\hline
    \end{tabular}
\end{table*}

\section{Results}

\subsection{Assessment}

To assess the quality of our course's labs, students completed an anonymous feedback form. Students answered a number of questions on a 7-point Likert scale, anchored by Strongly Disagree (1) and Strongly Agree (7). Table II shows the results of this survey for each Lab from the Fall 2025 academic semester. Additionally, to assess whether we reached our educational objectives set at the beginning of the course, students completed a survey before and after the course asking a number of questions about RF systems and wireless communications. This survey asked their self-efficacy in explaining a number of concepts regarding RF systems and wireless communications. Each response item asked students to rate their ability on a scale of 1 to 10. Students completed this survey at the beginning and end of the course, and these results are illustrated in Fig. 9.

			

\begin{figure}[h!]
    \centering
    \includegraphics[width=\linewidth]{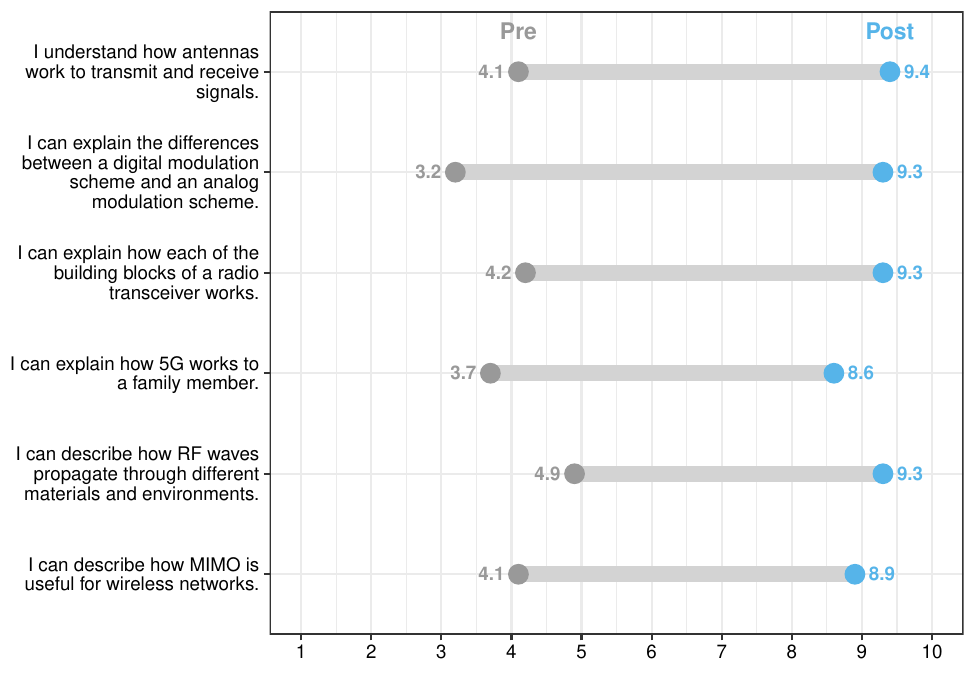}
    \caption{Results from pre-course (\textit{n=10}) and post-course (\textit{n=7}) self-efficacy surveys. Number by each point is the mean of responses across students. Prompts for all questions started with ``I am confident that...'' All pre-post response differences were significant as measured by unpaired t-test (\emph{p} \textless 0.001 for all questions).}
    \label{fig:enter-label}
\end{figure}


\subsection{Cost}

Table III summarizes the total cost of the course materials. Considering a class size of 20 students with teams of 2, the total one-time expenditure for course hardware is \$8,838. This sum excludes shipping costs, the cost of access to MATLAB, and the recurring cost of PCB manufacturing. Note that open-source software such as openEMS can be used to substitute the use of MATLAB with Python to eliminate this cost \cite{openEMS}. In comparison, to teach a course of this caliber, similar VNA and SDR options often found in the fields of wireless communications and RF engineering research can cost over \$2,000 per SDR and \$10,000 per VNA, dramatically increasing the cost of operating the course as the number of students grows \cite{ettus2025} \cite{copper2025}. Therefore, Table V presents a most affordable means of teaching RF systems in a hands-on manner.

\begin{table}[h!]
    \centering
\caption{Estimate of the hardware cost for the course, \\ assuming 10 groups of students.}
\label{tab:my_label}
    \begin{tabular}{|c|c|}\hline
         Item& Cost
\\\hline
         LibreVNA (x10)& \$700 (x10)
\\\hline
         HackRF& \$350
\\\hline
         RTL-SDR (x10)& \$48 (x10)
\\\hline
         LoRa Featherwing Microcontrollers and Antennas (x10)& \$20 (x10)
\\\hline
         Raspberry Pi 4& \$110
\\\hline
         KrakenSDR& \$499
\\\hline
         Kraken Antenna Array& \$199
\\\hline
         Total& \$8,838\\ \hline
    \end{tabular}
\end{table}

\clearpage

\section{Discussion and Conclusions}

A course exploring the design and implementation of RF systems is proposed, providing students with access to low-cost equipment and tools to apply these concepts. The course budget is less than \$9,000. The lab exercises in the course were positively received by students. Through the adoption of low-cost and open-source electronics, RF engineering may become more accessible to students of all backgrounds.



\bibliographystyle{IEEEtran}

\newpage
\bibliography{references}

\end{document}